# Two-dimensional spectral interpretation of time-dependent absorption near laser-coupled resonances

**Alexander Blättermann, Christian Ott, Andreas Kaldun, Thomas Ding, and Thomas Pfeifer**[1]

Max Planck Institute for Nuclear Physics, Saupfercheckweg 1, Heidelberg (Germany)

Email: tpfeifer@mpi-hd.mpg.de

**We demonstrate a two-dimensional time-domain spectroscopy method to extract amplitude and phase modifications of excited atomic states caused by the interaction with ultrashort laser pulses. The technique is based on Fourier analysis of the absorption spectrum of perturbed polarization decay. An analytical description of the method reveals how amplitude and phase information can be directly obtained from measurements. We apply the method experimentally to the helium atom, which is excited by attosecond-pulsed extreme ultraviolet light, to characterize laser-induced couplings of doubly-excited states.**

Studying the interaction of light and matter is one of the most active and fundamental research fields in modern physics. Since a few years, laser-driven high-order harmonic generation (HHG) provides a spectrally coherent light source in the extreme ultraviolet (XUV) with unprecedented spectral and temporal properties [1-3]. Perhaps the most fascinating aspect of this method is the generation of attosecond light bursts permitting close to atomic-unit temporal resolution in time-resolved experiments [4,5]. Here, the electronic properties of atoms [6,7], molecules [8-10], as well as solids [11-13] can be investigated on their natural attosecond time scale. So far, most attosecond experiments involve the generation and measurement of photoelectrons or ions, or focus on the harmonic generation process itself. Recently, an all-optical technique has gained increased attention, namely the so-called attosecond transient absorption spectroscopy [14-19], as an extension of femtosecond XUV transient absorption in the gas phase [20,21]. As compared to photoelectron detection, one key advantage of transient absorption is its intrinsic sensitivity to bound-to-bound transitions of the investigated quantum system, which manifests itself by sharp spectroscopic signatures known as spectral line shapes [22]. Utilizing attosecond XUV pulses in combination with synchronized near-infrared (NIR) femtosecond fields, a general mechanism that allows for the conversion of symmetric Lorentzian into asymmetric Fano [23] line shapes has recently been understood and applied by laser controlling the time-dependent dipole-response function [24].

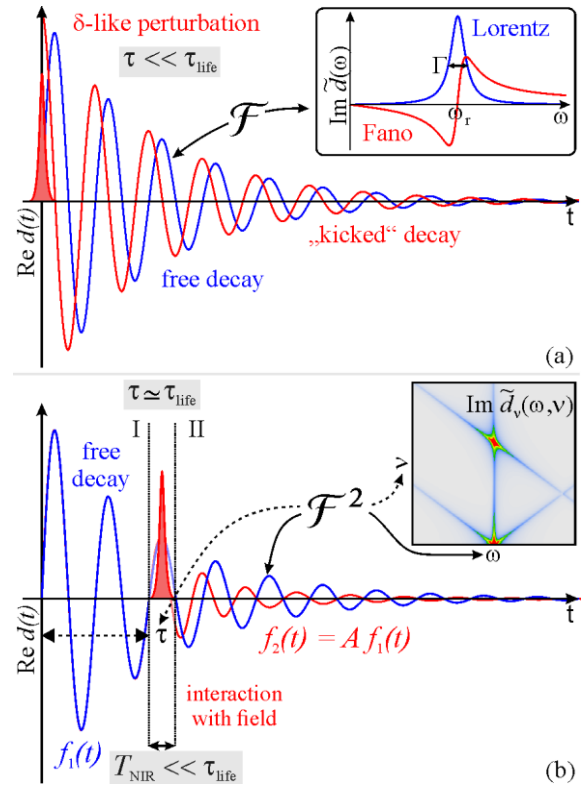

**Figure 1. Schematic illustration of the two-dimensional spectroscopy method**. (a) When the initial phase of the dipole response $d(t)$ of an excited atom is modified (indicated by the red pulse) at time $t \approx 0$, this leads to a characteristic change in the absorption spectrum $\sigma(\omega) \propto \mathrm{Im}\, \tilde{d}(\omega)$. (b) Generalization to phase or amplitude modifications at arbitrary time delay $\tau$ after excitation. Modifications occurring on time scales much shorter than the life time $\tau_{\mathrm{life}}$ of the state can be treated as an instantaneous amplitude and phase change of the dipole moment at time $\tau$. This is expressed by the complex quantity $A$, modifying the initial dipole response $f_1(t)$ to $f_2(t) = A\, f_1(t)$ at time $\tau$.

---

[1] Author to whom any correspondence should be addressed

Following the developments of femtosecond time-resolved spectroscopy, a technological revolution has occurred starting out from the first investigations on transient absorption in the mid 1980s [25-27]. One impressive example is the extension to two-dimensional electronic spectroscopy [28-31], a powerful technique that provides access to time-resolved reaction dynamics in complex systems. Moving towards higher photon energies in the XUV and soft-x-ray spectral region — also allowing for shorter pulse durations in the attosecond temporal region — theoretical work suggests to investigate the fundamental electronic couplings and natural attosecond dynamics in small quantum systems [32,33]. This dream can be expected to be implemented soon at next-generation free-electron-laser (FEL) light sources [34].

Here, we demonstrate a first experimental step towards two-dimensional spectroscopy with XUV attosecond-pulsed light. Based on the time-domain understanding of spectral line shapes, we create a complex-valued two-dimensional spectroscopic representation, which we obtain from the XUV response of laser-coupled autoionizing states [35-37]. Supported by an analytical framework, we show how such a representation can be interpreted and understood as an effective two-color (XUV and NIR) 3$^{\text{rd}}$-order nonlinear spectroscopy scheme. As in traditional two-dimensional spectroscopy the here introduced representation—though implemented in a conceptually different manner—reveals the coupling between coherently-excited quantum states. These first steps provide the key to understand the underlying dynamics of transiently coupled systems and lay the foundation towards other multidimensional nonlinear spectroscopy schemes with XUV and soft-x-ray light.

We start out by describing the temporal dipole response $d(t)$ of an isolated quantum-state resonance after a $\delta$-function-like excitation at $t = 0$, e.g. by an attosecond XUV pulse. Subject to an exponential decay rate $\Gamma$, and oscillating at a resonance frequency $\omega_r$, it is given by

$$d(t > 0) \propto -\mathrm{i}e^{\mathrm{i}\omega_r t - \frac{\Gamma}{2}t}. \tag{1}$$

The associated spectral line shape is obtained after Fourier transforming the dipole response, and evaluating its imaginary part given by

$$\begin{aligned}\operatorname{Im}\tilde{d}(\omega) &= \operatorname{Im}\left[\int_{-\infty}^{\infty} d(t)\mathrm{e}^{-\mathrm{i}\omega t}\mathrm{d}t\right] \\ &\propto \operatorname{Im}\left[\frac{1}{\mathrm{i}(\omega_r - \omega) + \Gamma/2}\right].\end{aligned} \tag{2}$$

This is a Lorentzian profile centered at $\omega_r$ with its full width at half maximum equal to the decay rate $\Gamma$ of the state [see figure 1 (a)]. For a dilute medium this quantity can be related to the experimentally measured absorption cross section [16,17,22,27] given by

$$\sigma(\omega) \propto \omega \operatorname{Im}\left[\tilde{d}(\omega)/\tilde{E}_{\text{XUV}}(\omega)\right]. \tag{3}$$

Here, $\tilde{E}_{\text{XUV}}(\omega)$ represents the complex spectrum of the excitation pulse, which is a constant for the case of excitation with a $\delta$-pulse. The linear increase of $\omega$ can also be safely neglected for the common case of $\Gamma/2 \ll \omega_r$ and in the absence of close-lying neighboring resonances. $\operatorname{Im}\tilde{d}(\omega)$ is therefore the observable quantity in a transient absorption spectrum for the short-pulse (impulsive) excitation case.

The effective initial phase of the dipole response can be controlled by means of an ultrashort laser pulse (e.g. NIR), which impulsively interacts with the electron in the upper radiating state at time $t \approx 0$, right after its population by the XUV attosecond pulse, as demonstrated in Ref. [24]. In the subsequently measured spectrum this leads to substantial modifications of the observed line shape. Controlling the phase with a tunable laser-field strength, Lorentzian line shapes can be converted into Fano-like line shapes, and absorption can be turned into transparency or even gain.

We now extend this framework to the control at arbitrary time delay $\tau$ between the excitation attosecond XUV and the coupling NIR laser pulse. The spectral line shape $\operatorname{Im}\tilde{d}(\omega)$ corresponding to such a perturbed polarization decay [38,39] is then parameterized by $\tau$, thus explicitly requiring a two-dimensional representation in $\omega$ and $\tau$.

The key idea is illustrated in figure 1 (b). In the temporal region I after the excitation by an attosecond XUV pulse (excitation at time $t = 0$; not shown) the system starts to radiate dipole emission, determined by its natural dynamics, in a field-free environment. Within the time window of duration $T_{\text{NIR}}$ centered at $\tau$ the system then interacts with an additional NIR laser field. Afterwards, in temporal region II the system again evolves in a field-free environment for the remainder of its natural life time $\tau_\text{life}$. Such a description is a general approach to describe the interaction of a system with a sequence of laser pulses of much shorter duration than the system's natural dynamics (*e.g.* the decay). In order to arrive at analytical expressions, and to keep track of the physically relevant features, we will restrict our discussion to the limiting case of $T_\text{NIR} \to 0$, thus treating both XUV excitation and NIR coupling as $\delta$-function-like inter-actions. This approximation is justified when the duration $T_\text{NIR}$ of the laser pulses is negligible compared to the life time $\tau_\text{life}$ of the system. For this case we can write the time-dependent dipole response in its most general form as

$$d_\tau(t, \tau) \propto \begin{cases} 0 & t < 0 \\ f_1(t) & 0 < t < \tau \\ A\, f_1(t) & t > \tau \end{cases} \quad (4)$$

where $f_1(t)$ describes the system's natural response, *e.g.* $f_1(t) = -\text{i} e^{\text{i}\omega_\text{r} t - \frac{\Gamma}{2} t}$ for an isolated resonance of a two-level system. $A$ is a complex-valued number which accounts in all generality for light-induced phase and amplitude modifications of the system. In general, $A$ will be a function of the time delay $\tau$: $A = A(\tau)$; with $A(\tau < 0) = 1$. To obtain the absorption spectrum $\text{Im}\,\tilde{d}_\tau(\omega, \tau)$, we evaluate the Fourier transform of $d_\tau(t, \tau)$:

$$\tilde{d}_\tau(\omega, \tau) = \int_{-\infty}^{\infty} d(t) e^{-\text{i}\omega t} \text{d}t$$
$$\propto \int_0^\tau f_1(t) e^{-\text{i}\omega t} \text{d}t + A(\tau) \int_\tau^\infty f_1(t) e^{-\text{i}\omega t} \text{d}t. \quad (5)$$

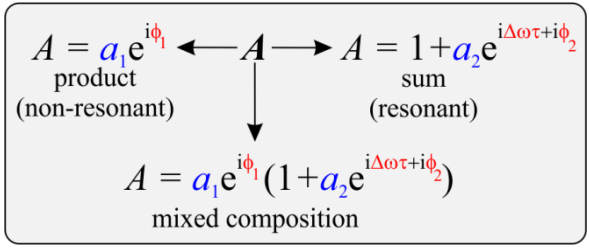

**Figure 2. The structure of the complex $A$, modifying the dipole response of a given quantum state in amplitude and phase.** In general, $A$ can be written as a product or a sum of complex terms. The product form is suitable to describe non-resonant processes like tunneling or non-resonant transient level shifts (e.g. AC Stark shift). The sum form applies for resonant processes, *e.g.* the perturbative coupling of states for small values of $a_2$, where $\Delta\omega$ is the resonance frequency of the process. In general, the phase and the magnitude can be static or a function of the time delay $\tau$ and other parameters, *e.g.* laser intensity, wavelength, or polarization. By constructing a composition of these basic building blocks of dipole control, more complex interactions can be described and understood.

Assuming free decay, the Fourier transform is analytically given by

$$\tilde{d}_\tau(\omega, \tau) \propto -\text{i}\,\frac{1 - e^{\text{i}(\omega_\text{r} - \omega)\tau - \Gamma/2\tau}\bigl(1 - A(\tau)\bigr)}{\text{i}(\omega_\text{r} - \omega) - \Gamma/2}. \quad (6)$$

For convenience we introduce the frequency detuning $\delta = \omega_\text{r} - \omega$ and the dipole spectrum $\widetilde{D}_\tau(\delta, \tau) = \tilde{d}_\tau(\omega_\text{r} - \delta, \tau)$ as well as $\gamma = \Gamma/2$. Equation (6) then reduces to

$$\widetilde{D}_\tau(\delta, \tau) \propto -\text{i}\,\frac{1 - e^{\text{i}\delta\tau - \gamma\tau}\bigl(1 - A(\tau)\bigr)}{\text{i}\delta - \gamma}. \quad (7)$$

From equation (7) we already gain some insight into the spectral response of the system:

For $A(\tau) = 1$, *i.e.* in the absence of the NIR laser pulse, the well-known Lorentzian line shape of the freely decaying two-level system results and is independent of $\tau$. For the case of non-zero interaction $[A(\tau) \neq 1]$, we identify several $\tau$-dependent phase terms [note that $A(\tau)$ is generally complex-valued], which give rise to characteristic spectral interference patterns [see figure 3 (a), (e), and (i), explained in more details below].

In the following, we show that the Fourier representation ($\tau \to \nu$) of such a time-delay scan reveals amplitude and phase information on the laser-induced modification of the quantum state. The key quantity of interest is the complex Fourier transform of the time-delay scan

$$\tilde{d}_\nu(\omega, \nu) = \int_{-\infty}^{\infty} \text{Im}\bigl[\tilde{d}_\tau(\omega, \tau)\bigr] e^{-\text{i}\nu\tau} \text{d}\tau \quad (8)$$

or alternatively

$$\widetilde{D}_\nu(\delta, \nu) = \int_{-\infty}^{\infty} \text{Im}\bigl[\widetilde{D}_\tau(\delta, \tau)\bigr] e^{-\text{i}\nu\tau} \text{d}\tau \quad (9)$$

with respect to $\tau$. This quantity casts the time-delay-dependent absorption spectrum into a two dimensional absorption spectrum (2DAS). The two spectral dimensions of the 2DAS are the frequency range $\omega$ of an excitation pulse, and $\nu$ of a coupling pulse. In this work, we use an XUV pulse for excitation and an NIR pulse for coupling, but the approach is entirely general and not limited to these spectral regions.

To further understand the structure of the 2DAS it is necessary to evaluate equation (9) for specific $A(\tau)$. The form of $A(\tau)$ depends on the interaction process. Here, we restrict ourselves to two fundamental cases:

1) A constant amplitude and phase modification factor $A = a_1 \, e^{i\phi_1}$. This can be used to model nonresonant processes, for instance the AC stark shift or strong-field ionization, the latter also reducing the population of a state [40].

2) Additive time-delay dependent amplitude and phase modification: $A = 1 + a_2 \, e^{i\Delta\omega\tau + \phi_2}$. This modification applies for perturbative resonant coupling to another coherently excited quantum state, where $\hbar\Delta\omega$ is the energy difference between the two states. These examples are summarized in figure 2. The quantity $A$ thus forms a key ingredient in this dipole-control model, as we will refer to it below.

In the following, we discuss two realistic scenarios for the two fundamental interactions (non-resonant: case 1; resonant: case 2). We choose the energies and life times of the contributing states to match the situation of doubly-excited helium, for which we also present experimental results below. For case 1 the imaginary part of equation (7) reads

$$\text{Im } \widetilde{D}_{\tau,1}(\delta,\tau) \propto \frac{1}{\delta^2 + \gamma^2} \times \{\gamma + e^{-\gamma\tau} \\ \times \\ e^{i\delta\tau}(\gamma + i\delta)(a_1 e^{i\phi_1} - 1)\} + \text{c.c.} \quad (10)$$

For the trivial situation $a_1 = 1$ and $\phi_1 = 0$, i.e. $A = 1$, the oscillatory term $e^{i\delta\tau}$ vanishes, which results in a time-delay-independent Lorentzian profile. Otherwise, an explicit time-delay dependence arises, and a modification of the line shape due to spectral interference occurs. In case 2 we obtain:

$$\text{Im } \widetilde{D}_{\tau,2}(\delta,\tau) \propto \frac{1}{\delta^2 + \gamma^2} \times \{\gamma + e^{-\gamma\tau} \times \\ a_2 \, e^{i(\delta + \Delta\omega)\tau + i\phi_2}(\gamma + i\delta)\} + \text{c.c.} \quad (11)$$

For $a_2 = 0$, i.e. vanishing perturbation, equation (11) also yields a time-delay-independent Lorentzian line shape. By contrast, $a_2 \neq 0$ again results in a time-delay-dependent absorption spectrum. The Fourier transform [equation (9)] can now be evaluated analytically so that it is possible to identify the terms of interest in the final representation shown in figure 3. For case 1, substituting equation (10) into equation (9) yields the 2DAS

$$\widetilde{D}_{\nu,1}(\delta,\nu) \propto \frac{1}{\delta^2 + \gamma^2} \Big\{ 2\gamma \, \delta_D(\nu) \\ + \Big[\frac{a_1 e^{i\phi_1} - 1}{i(\delta - \nu) - \gamma}(\gamma + i\delta) \\ - \frac{a_1 \, e^{-i\phi_1} - 1}{i(\delta + \nu) + \gamma}(\gamma - i\delta)\Big]\Big\}, \quad (12)$$

whereas for case 2 the respective result is

$$\widetilde{D}_{\nu,2}(\delta,\nu) \propto \frac{1}{\delta^2 + \gamma^2} \Big\{ 2\gamma \, \delta_D(\nu) \\ + a_2 \Big[\frac{e^{i\phi_2}}{i(\delta + \Delta\omega - \nu) - \gamma}(\gamma + i\delta) \\ - \frac{e^{-i\phi_2}}{i(\delta + \Delta\omega + \nu) + \gamma}(\gamma - i\delta)\Big]\Big\}, \quad (13)$$

where $\delta_D(\nu)$ is the Dirac-$\delta$ function. Figure 3 (a), (e), and (i) show the time-delay scans, i.e. the absorption spectrum Im $\tilde{d}_\tau(\omega,\tau)$ as a function of the delay, for some examples of non-resonant (case 1) and resonant (case 2) coupling. The observed structures closely resemble many recent attosecond transient absorption studies involving perturbed-polarization decay [14,39-46]. Figure 3(a) reveals hyperbolic structures ($\delta \times \tau = $ const.) accompanied by spectral broadening for $\tau \to 0$. This is a result of spectral interference of the two temporal regions [I and II in figure 1 (b)]. A physical example is the depletion of the excited state by means of a strong-field ionizing NIR laser pulse as considered in reference [40]. These strong hyperbolic features are suppressed in the other two examples [figure 3(e) and (i)] of weak resonant coupling. By contrast, they exhibit fast $\tau$-dependent rippling, which is due to quantum interference of the two coupled states, and thus characteristic for resonant coupling. The beating frequency $\Delta\omega$ is given by the energy difference of the states. According to the $e^{i(\delta + \Delta\omega)\tau}$ term in equation (11) the beating frequency along $\tau$ is modified by $\delta$, such that only at the resonance position ($\delta = 0$) the beating frequency exactly equals the energy spacing of the states. A result of this effect is a tilting of the ripples with increasing $\tau$.

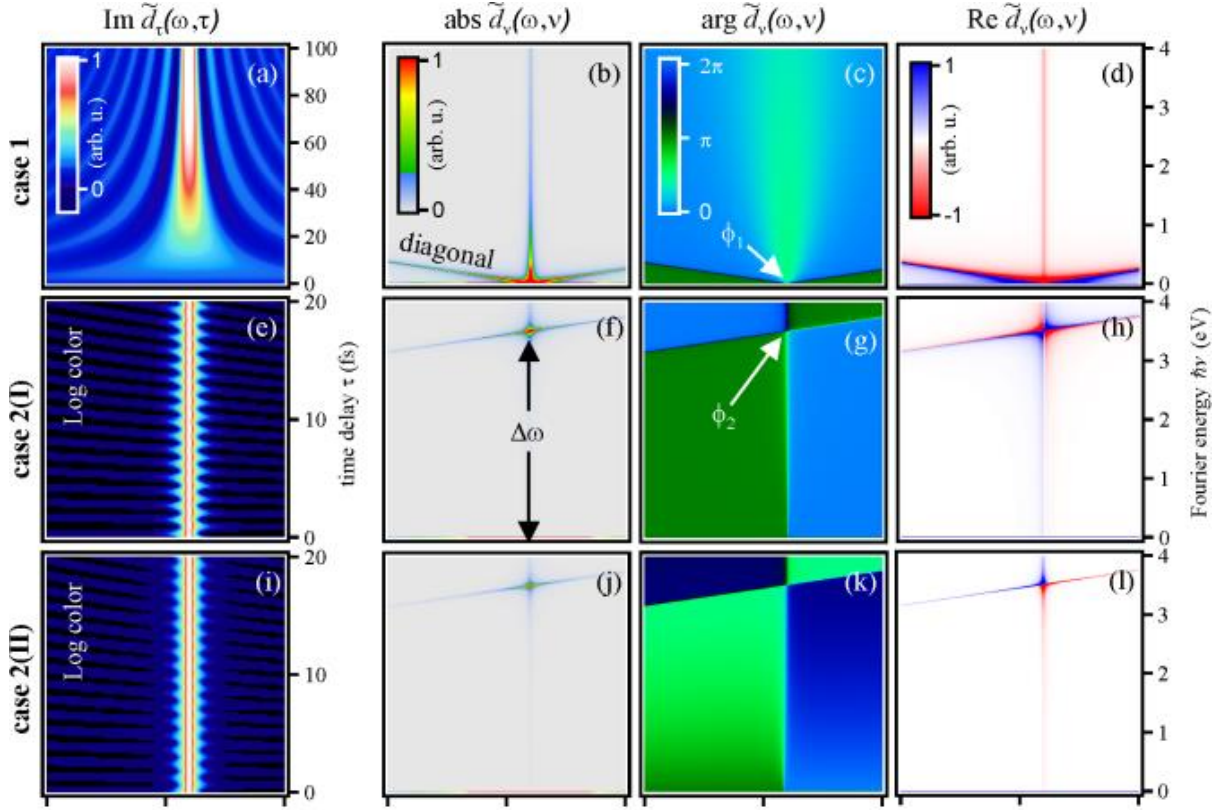

**Figure 3. Structural building blocks of the XUV-2DAS** (Dipole-control model results). Top row shows case 1, a factor-like modification of an excited-state dipole moment at time-delay $\tau$ after excitation (case 1: $A = a_1\,e^{i\phi_1}$; $a_1 = 0, \phi_1 = 0$ ). The lower rows show two examples of case 2, *i.e.* sum-like modification at $\tau$ (case 2: $A = 1 + a_2\,e^{i\Delta\omega\tau + \phi_2}$; (I): $a_2 = 0.1, \phi_2 = \pi, \hbar\Delta\omega = 3.51$ eV; (II): $a_2 = 0.05, \phi_2 = \pi/2, \hbar\Delta\omega = 3.51$ eV). We chose $\hbar\omega_r = 63.66$ eV as for the $sp_{23+}$ state in helium. The first column (a), (e) and (i) displays the time-delay scans Im $\tilde{d}_\tau(\omega,\tau)$ as directly observable in experiments. The second column [(b), (f), (j)] contains the magnitude of the 2DAS $\tilde{d}_\nu(\omega,\nu)$ of (a), (e), and (i). The three graphs exhibit diagonal and peak structures in direct correspondence to the functional form of $A$. Finally, the third [(c), (g), (k)] and fourth [(d), (h), (l)] column show phase and the real part of $\tilde{d}_\nu(\omega,\nu)$, respectively. The shape of Re $\tilde{d}_\nu(\omega,\nu)$ also allows for the retrieval of the laser-induced phase change.

Figures 3 (b), (f), (j), and (c), (g), (k), show magnitude and phase of the corresponding 2DAS $\tilde{d}_\nu(\omega,\nu)$, respectively. The last column [(d), (h), (l)] exhibits the real part of the 2DAS, which reveals different structural symmetries according to the phase modification. The characteristic structures of the 2DAS magnitude plots are peaks and diagonals of slope 1, which arise from equations (12) and (13) for the non-resonant case 1 and the resonant case 2, respectively. The two terms in square brackets represent complex Lorentzian profiles centered at the diagonal lines $\delta + \Delta\omega \mp \nu = 0$ in the two-dimensional representation for the resonant case 2. For the non-resonant case 1 the line is defined by $\delta \pm \nu = 0$ since the phase is not periodically modulated by $\tau$. This means that the vertical location of the peak at the resonance frequency, *i.e.* $\delta = 0$, directly reveals the modulation frequency of the laser-induced phase, which is 0 in case 1 and $\Delta\omega$ in case 2. Thus, the resonant coupling of states can be identified with peaks located at the energy difference of the coupled states in the 2DAS. The tilt of the diagonal is such that it points towards the energy of the coupled state on the $\delta$ axis. In contrast, non-resonant features of laser–atom interactions like strong-field ionization and ponderomotive or Stark shifts usually contribute near $\nu = 0$.

Finally, figures 3 (c), (g), (k) and (d), (h), (l) give access to the phase of the complex quantity $\tilde{d}_\nu(\omega,\nu)$. From this representation, the constant phase shift $\phi_{1,2}$ of the dipole moment right after interaction with the NIR laser pulse can be extracted. For the diagonals ($\delta + \Delta\omega \pm \nu = 0$) of the resonant interaction of case 2 equation (13) reduces to

$$\tilde{D}_{\nu,2}(\delta,\nu) \propto \frac{-a_2\,(\gamma \pm i\delta)}{\gamma(\delta^2 + \gamma^2)} e^{\pm i\phi_2}. \quad (14)$$

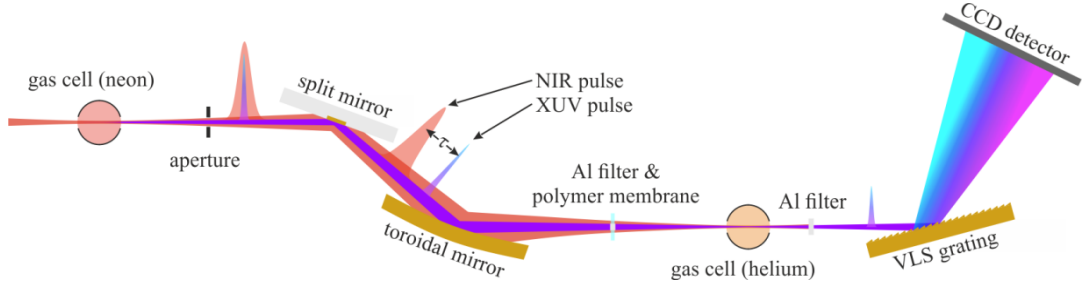

**Figure 4. Schematic view of the experimental setup.** The XUV light is generated by the NIR laser via HHG in a neon-filled gas cell. A split mirror is used to introduce a time delay $\tau$ between the co-propagating pulses, which are then refocused into the target gas cell containing helium. After reflection from a variable-line-space (VLS) grating the transmitted light is recorded with a CCD camera in flat-field geometry. An aperture and spectral filters are used to control the NIR intensity and to spatially separate the NIR from the XUV light, respectively.

At the resonance frequency $\delta = 0$ the formula further simplifies to $\widetilde{D}_{\nu,2}(0,\nu) = \frac{a_2}{\gamma^3} e^{i(\pm\phi_2 + \pi)}$. The phase is then extracted as

$$\pm\phi_2 = \arg \widetilde{D}_{\nu,2}(0,\nu) - \pi. \quad (15)$$

The same procedure can be applied to the non-resonant interaction of case 1. On the diagonal ($\delta \pm \nu = 0$) equation (12) evaluates to

$$\widetilde{D}_{\nu,1}(\delta,\nu) \propto \frac{-(a_1 e^{\pm i\phi_1} - 1)(\gamma \pm i\delta)}{\gamma(\delta^2 + \gamma^2)}. \quad (16)$$

From the argument of this expression the phase $\phi_1$ can also be retrieved.

With the basic structure of the 2DAS understood in terms of the general dipole-control model, we now apply the technique to a non-perturbative 3-level simulation of an absorption experiment in helium, and to a set of experimental data of the $sp_{23+}$ doubly excited state in helium. The experimental setup is shown schematically in figure 4. The $sp_{23+}$ state couples to the 2s2p state via a resonant two-NIR-photon transition (NIR laser intensity $2 \times 10^{12}$ W/cm$^2$, pulse duration $\sim 7$ fs, center wavelength 730 nm) via the intermediate $2p^2$ state. For additional experimental details see reference [24]. For the 3-level simulation we considered the states 2s2p ($\psi_1$), $2p^2$ ($\psi_2$), and $sp_{23+}$ ($\psi_3$). The XUV excitation of the states $\psi_1$ and $\psi_3$ is treated perturbatively ($\psi_2$ cannot be accessed from the ground state), whereas the laser-coupling of all excited states was treated as a strongly coupled 3-level system, for which we solved the time-dependent Schrödinger equation

$$i\frac{\partial}{\partial t}\begin{pmatrix}\psi_1\\\psi_2\\\psi_3\end{pmatrix} = \begin{pmatrix}H_1 & W_{12}(t) & W_{13}(t)\\W_{21}(t) & H_2 & W_{23}(t)\\W_{31}(t) & W_{32}(t) & H_3\end{pmatrix}\begin{pmatrix}\psi_1\\\psi_2\\\psi_3\end{pmatrix} \quad (17)$$

numerically. Here, $W_{ij}(t) = d_{ij} E_{\text{NIR}}(t)$ are the couplings in terms of the transition dipole moments $d_{ij}$ and the time-dependent laser field $E_{\text{NIR}}(t)$. To account for the autoionizing nature of these states, the diagonal matrix elements $H_j = E_j + i\gamma_j$ incorporate decay by means of their imaginary part. In addition, the wavefunctions are multiplied with the corresponding Fano phase factor [24] to account for the originally asymmetric Fano line shapes.

As discussed above, for case 2 of resonant coupling, we show in figure 5 how amplitude and phase information can be extracted from the measured and simulated data. The energy difference between the coupled states $sp_{23+}$ and 2s2p can be clearly identified from the vertical position of the diagonal structure at the resonance frequency $\omega = \omega_r$. It reads $\Delta E_{\text{ex}} = (3.5 \pm 0.1)$ eV in the experiment and $\Delta E_{\text{sim}} = (3.52 \pm 0.04)$ eV in the simulation, which agrees with the expected value $\Delta E = E_{sp_{23+}} - E_{2s2p} = 3.51$ eV obtained from spectra using synchrotron light sources [47,48].

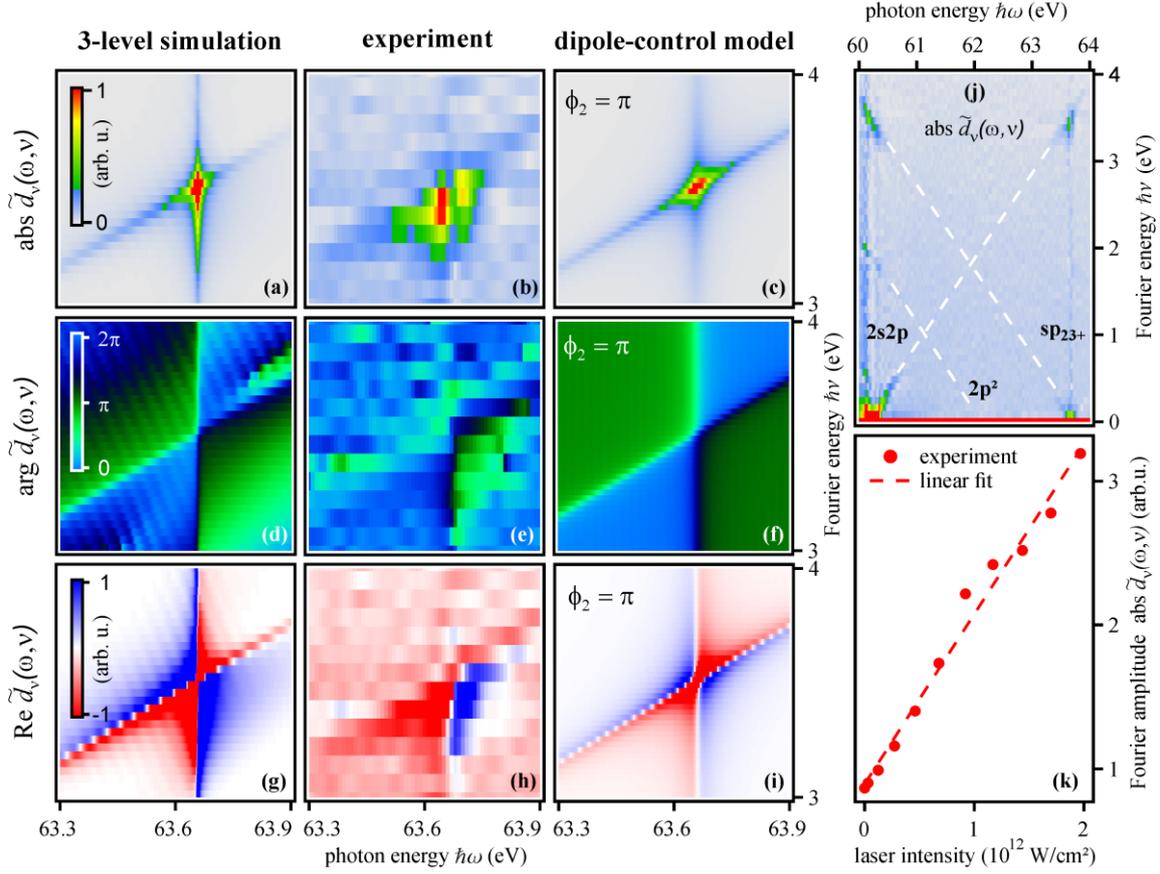

**Figure 5. Comparison of a 3-level non-perturbative simulation, experiment, and dipole-control model for amplitude and phase retrieval.** For (a) – (i) the rows indicate the shown quantity [magnitude, phase, and real part of $\tilde{d}_\nu(\omega,\nu)$] and columns indicate the data set [3-level simulation, experiment, and dipole-control model assuming $A(\tau) = 1 + a_2\, e^{i\Delta\omega\tau + \phi_2}$]. All plots show the coupling feature of the sp$_{23+}$ state. (j) shows an overview of the involved states in the 2d representation via abs $\tilde{d}_\nu(\omega,\nu)$ as observed in the experiment. (k) displays the integrated magnitude of the peak displayed in (b) versus the NIR laser intensity. It reveals the linear relation as expected from second-order perturbation theory. The linear fit does not intersect at the origin due to background in the 2DAS magnitude, which is caused by noise of the experimental data.

At this resonance frequency we read the phase jump $\phi_2$ directly from the phase plot, which yields for the simulation $\phi_{2,\text{sim}} = (\pi - 0.1 \pm 0.2)$ rad and for the experiment $\phi_{2,\text{ex}} = (\pi + 0.2 \pm 0.2)$ rad. In addition to directly reading out the phase we can compare the structure of the phase and the real part of the 2DAS with the model data calculated for a certain phase $\phi_2$. In figure 5 we show the result for $\phi_2 = \pi$, in good agreement with the experimental data. It is worth mentioning that the phase jump $\phi_{1,2}$ depends critically on the definition and the exact knowledge of the relative timing of excitation and coupling pulses. This is of particular importance when comparing different sets of data, e.g. experiment and simulation, where a common time basis ('time zero') has to be established on a sub-fs scale. To circumvent this issue, data with a common definition of time zero is required, which then enables direct access to time-dependent phase shifts.

Finally, figure 5 (k) shows the integrated magnitude of the Fourier peak versus the coupling laser intensity. For moderate intensities the relation is linear as expected from perturbation theory: For the two transition steps 2s2p →2p$^2$ (1→2) and 2p$^2$ → sp$_{23+}$ (2→3) the dipole coupling terms are $E_{\text{NIR}}d_{12}$ and $E_{\text{NIR}}d_{23}$. Thus, the amplitude $a_2$ (see figure 2) is proportional to $E_{\text{NIR}}d_{12}\, E_{\text{NIR}}d_{23} \propto I_{\text{NIR}}d_{12}d_{23}$.

In summary, we introduced a general framework to understand resonant and non-resonant time-delay-dependent dipole dynamics driven by short pulsed laser fields. In a two-dimensional representation (2DAS) resonant and non-resonant interaction pathways can be separated and analyzed. Furthermore, this method allows for the extraction of amplitude and phase modifications of atomic dipole moments of coupled states interacting with strong pulsed laser light. The analytical description provides physical insight into commonly observed features in transient absorption spectroscopy of perturbed polarization decay. As a first application, we experimentally observed laser-induced coupling of quantum states in the benchmark system of helium using ultrashort laser pulses. The presented understanding allows for a qualitative and quantitative comparison between theory and experiment, and thus proves the validity and power of the introduced 2d spectroscopy method. In the future the technique can be applied to study multi-state coupling processes, where the Fourier analysis allows to disentangle the individual couplings between states. It may thus also be helpful for studying more complex systems such as many-electron atoms and molecules. By adding a precise *in-situ* intensity measurement and characterization of the coupling laser it would even become possible to quantify the coupling dipole-matrix elements. The two-dimensional method introduced here allows to separate the contributing coupling pathways in a manner similar to—though qualitatively different from—traditional 2d spectroscopy [28], but allows for implementation in the XUV domain. The extension into the soft- and hard-x-ray spectral region at Free-Electron Lasers (FELs) appears feasible in the near future.

Acknowledgements:

Financial support from the Max Planck Research Group program and the Deutsche Forschungsgemeinschaft (grant no. PF 790/1-1) is gratefully acknowledged.